\documentclass[pra,twocolumn,floats,showpacs]{revtex4}
\usepackage{graphicx}
\usepackage{epsfig}
\usepackage{amssymb}
\usepackage{graphicx}
\usepackage{epsfig}
\usepackage{amssymb}
\newcommand{\gqH}{g_{\rm qH}}

\newcommand{\pr}{Phys.\ Rev.\ }

\newcommand{\be}{\begin{equation}}
\newcommand{\ee}{\end{equation}}
\renewcommand{\vec}[1]{{\mathbf #1}}

\begin{document}

\title{Quantum Hall states of atomic Bose gases: \\
       density profiles in single-layer and multi-layer geometries}

\author{N.R.~Cooper}
\affiliation{Theory of Condensed Matter Group, Cavendish Laboratory,\\
Madingley Road, Cambridge CB3 0HE, United Kingdom}

\author{F.J.M. van Lankvelt}
\affiliation{Rudolf Peierls Centre for Theoretical Physics,\\
1 Keble Road, Oxford OX1 3NP, United Kingdom}

\author{J.W.~Reijnders and K.~Schoutens}
\affiliation{Institute for Theoretical Physics, University of Amsterdam,\\
Valckenierstraat 65, 1018 XE~~Amsterdam, the Netherlands}



\date{28 July, 2005}


\begin{abstract}

\noindent
We describe the density profiles of confined atomic Bose gases in the
high-rotation limit, in single-layer and 
multi-layer geometries. 
We show that, in a local density approximation, 
the density in a single layer shows a landscape of quantized steps due 
to the formation of incompressible liquids, which are analogous 
to fractional quantum Hall liquids for a two-dimensional electron gas 
in a strong magnetic field.  
In a multi-layered set-up we find different phases, depending on
the strength of the inter-layer tunneling $t$. We discuss the situation
where a vortex lattice in the three-dimensional condensate (at large 
tunneling) undergoes quantum melting at a critical tunneling $t_{c_1}$.
For tunneling well below $t_{c_1}$ one expects weakly coupled or isolated 
layers, each exhibiting a landscape of quantum Hall liquids. After 
expansion, this gives a radial density distribution 
with characteristic features (cusps) that provide experimental 
signatures of the quantum Hall liquids.
\end{abstract}

\maketitle
\vspace{0.1in}
\section{Introduction}

Among the fascinating developments in the field of quantum
gases is the possibility to study correlated states of matter
in a setting that is entirely different from the traditional
setting of electrons in a solid state environment. A prime
example are fractional quantum Hall states, which are expected
when trapped atoms (bosons or fermions) are made to rotate at
ultra-high angular momentum \cite{WGS,CW,CWG,RJ,HM,PZC,RvLSR}.
While there has been steady progress in achieving high
angular momentum \cite{ENSexp,JILAexp}, the conditions for the
actual realization of these states have not yet been met.

The most direct experimental signature of electronic (fractional)
quantum Hall states, the quantization of the Hall conductance,
is not easily available for realizations of such states with
neutral atoms. It is thus important to investigate the experimental
signatures of atomic quantum Hall states. A number of largely 
complimentary proposals have been made
\begin{itemize}
\item
fractional (braid) statistics
\cite{PFCZ}
\item
loss of condensate fraction
\cite{SHM1}
\item
detection of gapless edge excitations
\cite{Ca}
\item
detection of density correlations in expansion image
\cite{RC}
\item
characteristic density profiles \cite {vLRS,Co}
\end{itemize}
In this paper, we work out the proposals put forward in
\cite{vLRS,Co,letterCvLRS} for the detection of atomic (fractional) 
quantum Hall states via characteristic density profiles in a suitably 
engineered experimental set up.

We first analyze the case of a single layer of rotating atoms, with
a weak confining potential. For this situation we find a characteristic 
plateau landscape for the density, with steps at sharply quantized 
values of the density (see Figs.~\ref{fig:steps} and \ref{fig:steps2}),
which are direct consequences of the incompressibility of the quantum Hall states.

After that we consider a more involved situation, where a condensate
that is rotating around the $z$-axis is exposed to an optical
lattice potential in the $z$-direction. This leads to a multi-layer geometry, 
with an inter-layer tunneling whose strength is set by the strength
of the optical lattice potential. This approach is presently pursued
by a number of experimental groups. With an appropriate choice of
parameters, one expects a regime that can be characterized as {\it a
stack of weakly coupled layers that each display a landscape of
quantum Hall states}. We propose to probe this state after expansion
in both the axial and radial directions. As the expansion of each layer in the 
axial direction is very fast, it gives rise to a density 
profile as a function of the radial distance $r$ that is an average over the
radial density profiles of the many layers. Despite the fact that the density
profiles in these layers differ from each other (owing to the
differing numbers of particles in the layers), we show that the resulting
averaged density  
profile has characteristic features (cusps), which are remnants of 
the steps in the individual layers before expansion (see 
Figs.~\ref{fig:profile1}, \ref{fig:profile2}). 
We propose these cusps as possible experimental 
fingerprints of the atomic quantum Hall states.

Our presentation is organized as follows. In section \ref{sec:single} 
we discuss the case of a single layer, and present the expected landscape 
of quantized plateaus. In section \ref{sec:multilayer} we discuss the 
multi-layer set-up: section \ref{sec:melting} discusses the melting of 
the vortex lattice in the multi-layer case and section \ref{sec:phases} 
discusses the quantum phases expected after the melting. 
In section \ref{sec:expansion} we discuss the expansion after switching 
off the optical lattice and trapping potential and present the density profiles that are expected. 
Throughout the paper we indicate typical values of the experimental 
parameters that are involved, and the conditions needed to justify the 
approximations that we make.

\section{Single-layer geometry}
\label{sec:single}

\subsection{Experimental set-up}
We consider a single cloud of rotating bosons, with harmonic confinement 
$\omega_\perp$ in the $x-y$ (in-plane) direction and 
$\omega_\parallel$ in the $z$ (out-of-plane) direction.
The characteristic length scales $\ell_{\perp}$ and $\ell_\parallel$ 
are
\be
\ell_{\perp,\parallel} = \sqrt{\hbar/(m\omega_{\perp,\parallel})}\ .
\ee
We assume that the rotation frequency $\omega$ is at or near the 
critical frequency $\omega_\perp$, and that parameters are such that
the system is in the two-dimensional (2D) regime. In this set-up the 
energy scale for atom-atom interactions is
\be
g_{\rm qH} = \frac{g}{(2\pi)^{3/2}\ell_\parallel \ell_\perp^2}
\label{eq:gqH-def}
\ee
with $g = \frac{4\pi \hbar^2 a_s}{m}$, where $a_s$ is the scattering
length. Assuming the values $\hbar \omega_\perp = 5$nK, $a_s=5$nm, 
and taking $\ell_\parallel= 50$nm for a layer defined by an optical 
lattice (see below), gives an energy scale $\gqH \simeq 0.5$nK.
Since typically $\gqH < \hbar \omega_\parallel, \hbar\omega_\perp$, 
in the study of quantum
Hall states, the single particle states can be restricted to the 
two-dimensional lowest Landau Level (LLL) \cite{WGS}. Clearly, the 
observation of quantum
Hall states (in single or multi-layer set-up) will require that
all energy scales in the experiment (tunneling, temperature, etc.)
are well below $\gqH$.

We mostly focus on one of the following two situations, which
we refer to as `quadratic' and `quartic' in-plane confinement.
\begin{description}
\item[quadratic confinement.]
Here we assume a rotation frequency $\omega$ slightly below the
critical frequency $\omega_\perp$, leaving a residual parabolic
potential in the rotating frame of reference $V_2(r)= {1 \over 2}
k_2 r^2$ with $k_2 \propto (\omega_\perp - \omega)$. 
\item[quartic confinement.]
This is for critical rotation $\omega=\omega_\perp$, but in the 
presence of an additional confining potential, which we take to 
be a quartic, $V_4(r) = {1 \over 4} k_4 r^4$ \cite{ENSexp}. 
\end{description}

\subsection{Phase separation in an external potential}
\label{sec:phasesep}

We refer to Refs.~\cite{WGS,CW,CWG,RJ,HM,PZC,RvLSR} for extensive
studies of the quantum liquids that form at high angular momentum,
after the point where the vortex lattice melts has been crossed.
Exact diagonalization studies on a disc, which represent a finite-size
gas in a harmonic confinement potential (for up to $N=10$ particles),
have revealed ``vortex liquid'' groundstates closely related to
fractional quantum Hall states\cite{WGS,CW}. Due to finite size
effects, the density profiles of these states show large fluctuations.
Features of some (but not all) of these states have been understood in
a picture invoking non-interacting composite
fermions\cite{CW}. Studies in edgeless geometries (sphere or torus)
have established the existence of homogeneous quantum Hall
liquids\cite{CWG,RJ}.  In these geometries the liquids have a 
uniform density over the entire system, so the calculations
represent regions of a trapped gas where the density is approximately
uniform.  Studies on edgeless geometries provide the most accurate
numerical data concerning the bulk properties of the quantum Hall
liquids that form at high rotation rate.

In this paper we focus on inhomogeneous quantum Hall liquids, with density
modulated by an external potential such as $V_2(r)$ or $V_4(r)$. The idea 
is the following. From the numerical studies cited earlier, we infer the
existence of stable, incompressible quantum liquids at certain specific
filling factors $\nu_i$ (the filling factor is defined below in terms of 
the ratio of the number densities of bosons and vortices), and we know the 
energy per particle $\epsilon_i$ in these liquids. If we now consider a 
given number of particles in a slowly varying external potential, we can 
patch together regions where the 
particles form specific quantum liquids, such as to minimize the total 
energy with the one global constraint of reproducing the total of number of 
particles in the trap.

For inhomogeneous systems, in which the potential $V({\bf r})$ varies
slowly in space compared to $\ell_\perp$, the density
distribution, $n({\bf r})$, (the number of atoms per unit area) averaged on scales large compared to
$\ell_\perp$, can be obtained by minimizing the functional
\be
  E = \int d^2 {\bf r} 
      \left( e[\nu({\bf r})] + V({\bf r}) n({\bf r}) - \mu n({\bf r}) \right) \ ,
\ee
where $e[\nu({\bf r})]$ is the interaction energy per unit area, which
is a function of the local filling factor $\nu({\bf r}) \equiv
n({\bf r})/n_0$ [with $n_0 = 1/(\pi\ell_\perp^2)$], and $\mu$ is the 
chemical potential (which controls the total number of particles).

That the interaction energy is purely local is a significant
simplification, arising from the fact that the interactions are
short ranged. Minimization of the functional then poses a purely
local problem. At each position, the density $n({\bf r})$ is that which
minimizes
\be
   e[\nu({\bf r})] - \mu_L n({\bf r})
\label{eqn:energy}
\ee
where the local chemical potential is
\be
   \mu_L({\bf r})  \equiv \mu - V({\bf r}) \ .
\label{eqn:muL}
\ee

For a vortex lattice (at large filling factor), the energy
density is
\be
  e[\nu] = n_0 b \nu^2  \times \gqH \ ,
\ee
where $b = 1.1596$ is the Abrikosov parameter for a triangular
lattice.  In this case, the dependence of density on local
chemical potential is simple:
\be
    n(r) =  n_0 \mu_L(r)/(2 b) \ ,
\ee
leading to a Thomas-Fermi profile \cite{FeBa} even in the 
LLL \cite{WaCo}.

For filling factors below a critical value $\nu_c$ 
(with $\nu_c$ on the order of 10 \cite{CWG,SHM2}) 
the vortex lattice is replaced by groundstates
that include incompressible quantum Hall fluids. The
energy density is then a complicated function of $\nu$, containing
cusps at the incompressible groundstates. The presence of these
cusps gives rise to a step-like dependence of $n$ on $\mu_L$.
Taken with Eq. (\ref{eqn:muL}), this step-like dependence of $n$
on $\mu_L$ becomes a step-like dependence in space for a confined
system (for which $V({\bf r})$ is not constant in space).

We remark that non-interacting fermions in a rapid rotation
regime, and subject to a slowly varying potential, will display
density profiles similar to the ones we discuss here \cite{HC}.
In that case, the result follows from the Landau level structure
for non-interacting fermions, while all results discussed in this
paper are due to interaction effects. The analogy between the two 
cases is strengthened by the `composite fermion' formulation of some
of our bosonic liquids. This analogy remains incomplete, however,
as some of the features that we see (such as the competition 
between composite fermion states and the paired Moore-Read state)
cannot be captured by a model of non-interacting composite fermions.

\subsection{Example: two liquids}
As a simple example, we consider the possibility of two competing
incompressible states: $\nu_1 = 1/2$ (the Laughlin state, for which 
the interaction energy for contact interactions vanishes, 
$\epsilon_1=0$\cite{WGS}), and $\nu_2=2/3$ for which the interaction energy
per particle is $\epsilon_2$. These two liquids are the $p=1,2$
members of a series at $\nu_p=p/(p+1)$, which can be
understood as integer quantum Hall liquids of composite fermions
each consisting of a boson with a single vortex attached
\cite{CW,RJ}.  Since the system is bounded, we must also consider
the case of no particles, $\nu=0$. The energy per unit area takes
the values $e[\nu=0]=e[\nu=1/2] = 0$, and $e[\nu=2/3] = (2/3) n_0
\epsilon_2$. Minimizing Eq. (\ref{eqn:energy}) with respect to
these three possible values of the density, $n = \nu n_0$, we
find:
\begin{eqnarray}
& \nu = 0             & \qquad (\mu < 0)
\nonumber\\
& \nu = 1/2           & \qquad (0< \mu < \mu_c)
\nonumber\\
& \nu = 2/3           & \qquad (\mu_c < \mu) \ .
\end{eqnarray}
with $\mu_c = 4 \epsilon_2$.

Let us take a harmonic confining potential $V(r) = {1 \over 2}
k_2 r^2$, such that $ \mu_L(r) = \mu - {1 \over 2} k_2 r^2$. The
$\nu=1/2$ state forms a disc that extends out to $\mu_L=0$
(beyond this $\mu_L<0$ so $\nu=0$), to a radius $r_1 =
\sqrt{2\mu/k_2}$. We further see that there is a critical value
of chemical potential, $\mu_c = 4 \epsilon_2$, at which the
$\nu=2/3$ state will first appear in the center of the trap
(where $\mu_L$ is maximum). Evaluating the total number of atoms
in the $\nu=1/2$ disc at $\mu_c$, we find the critical number
\be
  N_c =  (1/2) n_0 \pi r_1^2  = 2 \epsilon_2/\lambda_2
\ee
with $\lambda_2 = {1 \over 2} k_2 \ell_\perp^2$. Above this critical
value, the disc of $\nu=2/3$ has a radius
$r_2=\sqrt{2(\mu-\mu_c)/k_2}$.

Thus, as the number of particles increases, at first they will
form a disc of uniform density $1/2 \, n_0$ the radius of which
increases with increasing $N$ ($\mu$). Once $N$ reaches $N_c$
(or $\mu$ reaches $\mu_c$) a second, inner, disc of density $2/3 \,
n_0$ will form, and start to expand with increasing $N$ (or
$\mu$). In terms of $N$ and $N_c$ the two steps for $N>N_c$ are
located at radii
\be
  r_2/\ell_\perp = \sqrt{{3 \over 2} (N-N_c)} \ ,
  \quad
  r_1/\ell_\perp = \sqrt{{3 \over 2} N + {1 \over 2} N_c} \ .
\ee
The energy as a function of $N$ takes the form
\be
  E(N) = \lambda_2 \left(
       N^2 - {1 \over 4} \Theta(N-N_c) (N-N_c)^2 \right)
\label{EN-12}
\ee
We note that the second derivative of $E(N)$ is discontinuous at
$N=N_c$.

Note that, although we have only considered the possibilities
that $\nu=0, 1/2, 2/3$ (and we have not allowed for intermediate
values), we do not expect intermediate densities to appear in the
density distribution.  Since $e[\nu]=0$ for $\nu<1/2$, values of
$\nu$ between 0 and 1/2 are only possible precisely at $\mu_L=0$
and hence have no finite extent in a generic inhomogeneous
system. For the range, $1/2< \nu < 2/3$, numerical studies
indicate that $e[\nu]$ lies above the straight line joining the
end-points $e[\nu=1/2] = 0$ and $e[\nu=2/3]= 2/3 \, n_0
\epsilon_2$. Therefore, intermediate filling factors are unstable to phase
separation into $\nu=1/2$ and $\nu=2/3$.

A slightly different perspective on the formation of two
spatially separated liquids is the following. If one imagines
adding more and more particles to a $\nu=1/2$ Laughlin droplet in
a (radially symmetric) potential, one will eventually reach the
point where the energy of an additional particle at the edge of
the Laughlin droplet exceeds the energy gap $\Delta_1$
to create a single particle in the bulk Laughlin liquid at the 
center of the droplet. The corresponding critical value
for $N$ is $\widetilde{N}_c = \Delta_1 / (2 \lambda_2)$.
Using numerical results for $\Delta_1$ and $\epsilon_2$ (as
given, for example in \cite{RJ}) $\widetilde{N}_c$ is found to be
slightly higher than $N_c$, in agreement with the fact that
a $\nu=1/2$ system with additional quasi-particles will lower its
energy by phase separating into a central $\nu=2/3$ liquid and an
outer ring of $\nu=1/2$ liquid.

Repeating the analysis of the co-existing $\nu=1/2$ and $\nu=2/3$ liquids
in a quartic potential $V_4(r)={1 \over 4} k_4 r^4$ leads to a critical
$N$ of
\be
N_c = \sqrt{\epsilon_2 \over \lambda_4} \ ,
\ee
with $\lambda_4 = {1 \over 4}k_4 \ell_\perp^2$, and with inner and
outer edges at
\begin{eqnarray}
(r_2/\ell_\perp)^2 &=& {3 \over 4}(\sqrt{9N^2-8N_c^2}-N) \ ,
\nonumber\\
(r_1/\ell_\perp)^2 &=& {9 \over 4} N - {1 \over 4} \sqrt{9N^2-8N_c^2} \ .
\end{eqnarray}

The condition that $V(r)$ be slowly varying, phrased as 
${dV(r) \over dr}\ell_\perp \ll \gqH$, is equivalent to demanding
that $r_i / \ell_\perp \gg 1$ for all $i$. For this to be possible,
a total number of particles on the order of $N=100$ or more is required.
This puts the quantized density profiles out of the reach of exact 
diagonalization studies.

\subsection{More liquids}
Including in the analysis more quantum liquids,
at filling factors $\nu_1<\nu_2<...$
leads to a sequence of critical values $N=N_c^{(k)}$, marking the
start of the formation of a central region of a quantum Hall
fluid at $\nu=\nu_k$.

\begin{figure}
\epsfig{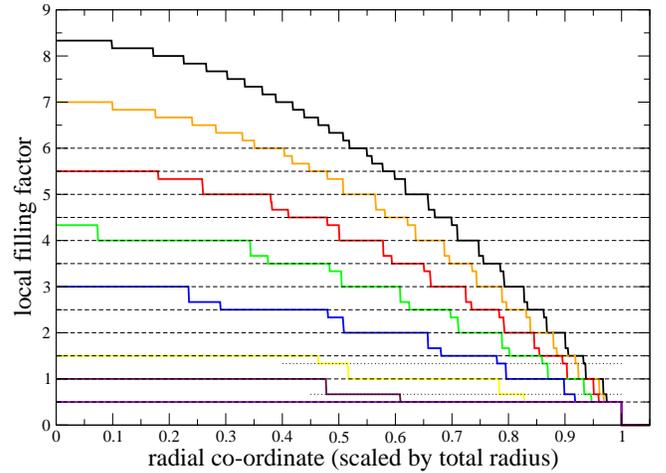} \caption{Density profiles
in a single layer, at subcritical rotation, parametrized by density
at the center of the trap. The results are for
an energy functional obtained from exact diagonalizations of a system with
$N_V=6$ states in toroidal geometry
(aspect ratio $a/b=\sqrt{3}/2$).}
\label{fig:steps}
\end{figure}

To illustrate the step-like profiles for higher filling factors, we use
the interaction energy function $e[\nu]$ calculated by exact
diagonalization studies on a torus \cite{CWG}. In the graphs presented in
Fig.~\ref{fig:steps}, a confining potential $V(r) = {1 \over 2} k_2 r^2$
is assumed.  The energy functional used is for system size of $N_V= 6$
single-particle states on a torus [for which the numerical results extend
up to $\nu=25/3$].

\begin{figure}
\epsfig{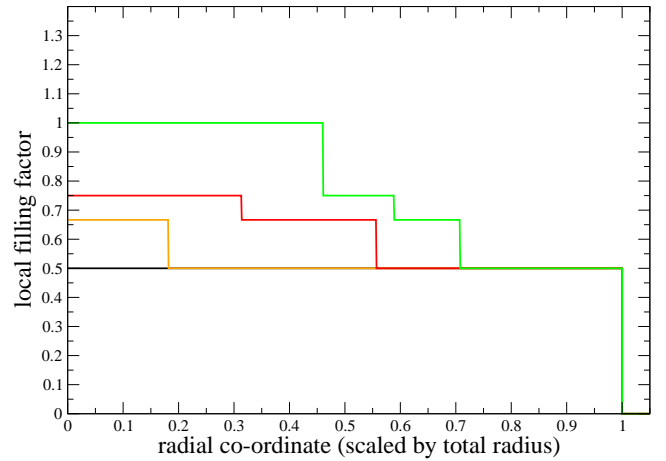} \caption{Density profiles
in a single layer, for cases with maximum central densities
of up to $\nu=1$. The energy functional is for $N_V=12$ states in toroidal
geometry
(aspect ratio $a/b=1$).The quantum liquid at $\nu=1$ is the Moore-Read
paired state and the $\nu=1/2$ state is the Laughlin liquid. In between,
a number of composite fermion states (at $\nu=2/3,3/4$) are observed.}
\label{fig:steps2}
\end{figure}

For filling factors larger than $\nu_c \simeq 6$, the groundstate is a
compressible vortex lattice \cite{CWG}. The steps seen in the density
distribution in that regime are an artifact of the minimum change of
filling factor, $\Delta \nu = 1/6$, allowed by the numerics on a finite
system with $N_V=6$. In this regime, the density profile will be smooth
(we emphasize again that our results apply only on scales large compared
to $\ell_\perp$, so do not describe the short-range density modulations
associated with the unit cell of a triangular vortex lattice).

For filling factors less than $\nu_c \sim 6$, these calculations show
plateaus appearing at the densities of a set of incompressible quantum
Hall fluids, including the sequence of Moore-Read and Read-Rezayi states
at $\nu=k/2$ (black dashed lines) as well as some of the CF sequence $\nu
= p/(p+1)$ (black dotted lines).  We note that finite size effects in
these results for $N_V=6$ can affect the widths and positions of the
plateaus appearing in Fig.~\ref{fig:steps}. This figure illustrates the
general feature of the formation of plateaus at the densities of
incompressible states. One interesting detail is the competition between
successive members of the CF series at $\nu={p}/{(p+1)}$, and the Moore-Read
(MR) liquid\cite{MR,CWG,RJ} at $\nu=1$. These states have very different
correlations, and one does not expect the higher CF states to be stable
against the pairing instability that is at the basis of the formation of
the MR state. This competition is highlighted in Fig.~\ref{fig:steps2},
which shows the density profiles for systems with maximum central
densities of up to $\nu=1$ calculated for a system with $N_V=12$ vortices
(which, unlike $N_V=6$, also permits the appearance of $\nu=3/4$, and for
which finite size effects on the plateau widths are much reduced). Note
that there is no appearance of an incompressible state at $\nu=5/6$. This
indicates that, at least within this small system calculation, the
$\nu=5/6$ state appears to be unstable to phase separation into the
$\nu=3/4$ and $\nu=1$ states.

\section{Multi-layer geometry}
\label{sec:multilayer}

\subsection{Experimental set-up}

We now investigate the situation where an optical lattice in the $z$
direction is imposed on a cigar shaped cloud of atoms rotating around
the $z$-axis. If the optical potential is sufficiently deep, this will
define a stack of parallel planes with (weak) tunneling $t$ between
the planes.  The idea is that parameters can be chosen such that,
while the original cloud is in a mean field regime with total filling factor
$\nu > \nu_c$ such that it would display a
vortex lattice, the individual layers defined by the optical lattice
can be in a quantum regime, so that the entire configuration becomes a
stack of weakly coupled quantum liquids. We will assume the presence
of a chemical potential $\mu(z) = \mu_0 - \mu_2 (z/d)^2$ (with $d$ the
distance between the layers), which will induce a slow modulation in
the number of particles per layer. One may view a set-up like this as 
a road towards isolating a single layer in the quantum Hall regime, or 
to perform diagnostics on the entire multi-layer geometry.

Reaching the regime where quantum Hall states can form requires a
number of conditions. First of all, to avoid the formation of a
mean-field vortex lattice, the filling factor in each of the layers
should be sufficiently low (meaning high-enough total angular momentum
per particle) and the interlayer coupling $t$ should be sufficiently
small. The second condition is that the tunneling is well below the
gap expected for the quantum Hall states.

Concerning the first condition, a quick estimate is made as follows.
The critical filling factor $\nu_c$ for a single layer has been found
to be of the order $\nu_c\sim 10$. Assuming $N$ particles,
$N_V$ vortices and $N_L$ layers, quantum melting will be possible
if $N/(N_L N_V)<\nu_c$. Taking, for example, $N=5000$ particles 
and $N_V=100$ vortices (filling factor $\nu=50$), slicing up the condensate 
into, say, $N_L=50$ layers of 100 particles each would produce a filling 
factor per layer $\nu^\prime$ of order unity. For this choice of 
parameters, a gradual lowering of the tunneling amplitude causes a 
quantum melting of the vortex lattice at a critical $t=t_{c_1}$.  
For $t<t_{c_1}$ one expects a state with quantum liquids in each of 
the layers.

The quantum Hall gap that features in the second condition is set by 
the interactions between the atoms (as encoded in the scattering length 
$a_s$) and the length scales $\ell_\perp$ and $\ell_\parallel$ of the 
confinement in the $x-y$ and $z$ directions. (In the multi-layer case, 
$\ell_\parallel$ refers to the thickness of a single layer.)
A theoretical estimate\cite{CWG,RJ} gives
\be
\Delta
\simeq 0.1 \, 4\pi \gqH
\simeq {a_s \over \ell_\parallel} \hbar \omega_\perp \ .
\label{eq:gqH}
\ee
with $\gqH$ as in Eq.~(\ref{eq:gqH-def}). With the
parameters as given below Eq.~(\ref{eq:gqH-def}), the quantum Hall
gap is found to be in the order of 1~nK.

There are several possibilities for experimental detection of the 
formation of incompressible states in the multi-layer set-up. One
possibility, which we worked out in \cite{letterCvLRS}, is to focus 
on the total particle number per layer, {\it i.e.}\ on the density 
profile as a function of the layer co-ordinate $\rho(z)$. 
In this paper we focus on a detection scheme where one allows 
a rapid expansion of the system and then measures the density 
profile as a function of the radial coordinate $r$. 

\subsection{Quantum melting of the vortex lattice}
\label{sec:melting}
We now turn to the analysis of the quantum melting of the vortex lattice
in the multi-layer geometry. For this, we will assume that all
layers have identical axial and radial confinement, with a chemical
potential independent of $z$. If the filling factor per layer, 
$\nu^\prime= N/(N_L N_V)$, is well
below the critical filling factor $\nu_c$ for a single layer, isolated 
single layers will be in the regime where the vortex lattice has 
melted. This means that for this situation, a gradual lowering of
$t$ (by turning up the strength of the optical lattice potential),
will lead to a melting of the vortex lattice at a critical
tunneling $t=t_{c_1}$. Equivalently, there is a critical
layer filling factor $\nu_c^\prime(t)$ 
for given tunneling $t$. 
We will use two inequivalent techniques to find the value of 
$\nu^\prime_c(t)$.  

Due to the short-range nature of the interaction, the inter-layer 
interactions are much smaller than both the intra-layer interaction 
and the tunneling, mean-field theory then predicts that the
vortex lattices in different layers will lie on top of each other.
Note that this is very different from the quantum Hall multi-layer
setting in semiconductor devices, where the long-ranged Coulomb
interaction can dominate the tunneling.

We model the coupled layers by including a Josephson term:
\begin{eqnarray}
H&=&\sum_i H^{(0)}_i(\mu) - \sum_{\langle ij\rangle}
	\int\!d^2\vec r\,t(\psi^\dagger_i\psi_j
			 + \psi^\dagger_j\psi_i),
\end{eqnarray}
where we include tunneling only between neighboring layers, labeled by $i$.

\begin{figure}
\centerline{\epsfig{file=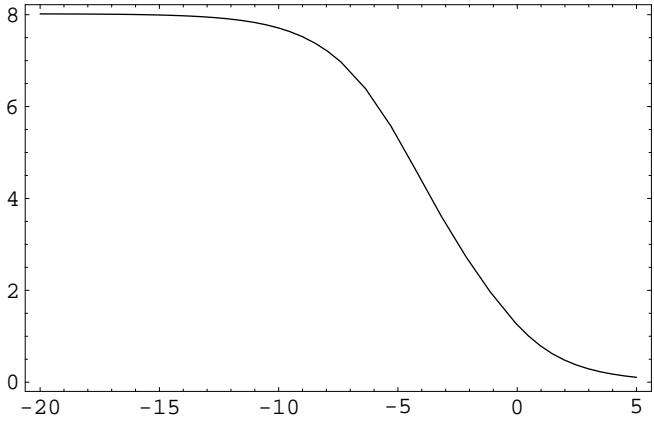,width=0.5\textwidth}}
\caption{Critical layer filling factor $\nu_c^\prime(t)$ as a function 
of $\ln(t/(\nu^\prime \gqH))$, obtained by applying a Lindemann criterion
to the vortex fluctuations (Eq.\ \ref{eq:lindemann}) in the limit 
$N_L\to\infty$, $N_V\to\infty$.}
\label{fig:lindemann}
\end{figure}

Our first approach for determining $\nu^\prime_c(t)$ is to extend 
the single layer effective theory of vortex fluctuations derived by 
Sinova {\it et al.} \cite{SHM2} to the multi-layer case. We neglect 
the gapped sound mode\cite{GB} and focus on the low-energy Tkachenko modes.
In terms of the deviations $\vec{u}^i(\vec r)$ of the vortex positions from
their equilibrium values, the Josephson coupling takes the following form:
\begin{equation}
\int_0^\infty\!\!\!d\tau\!\int\!\!d\vec q\,
        \sum_{\langle ij\rangle}\left[
	  \frac12t\frac{\alpha_1}{q^2}\left(
	    |u^i_L-u^j_L|^2 + |u^i_T-u^j_T|^2
	  \right)\right],
\end{equation}
with an additional (constant) contribution to the chemical potential.
In addition to the vortex positions, each layer has a phase
$\varphi_i$\@.  We assume that the phases of all layers are identical,
an approximation which is valid when $t\gg t_{c_2}$\@.

On distances smaller than the axial healing length (in units of the
optical lattice period $d$) $\xi_\parallel\approx \sqrt{t/\gqH \nu'}$,
all bosons are in the same axial state.  Therefore, they behave as a
single layer and the vortices are `rigid' on the scale of
$\xi_\parallel$.  In order to capture this effect, we  use
$\xi_\parallel$ as a short-distance cut-off in the axial direction.
The mean square displacement of a vortex is given by:
\begin{eqnarray}
\langle u^2\rangle &\approx&
    \frac1{2 N_L}\sum_k \int\!\!\frac{d^2\vec q}{(2\pi)^2}\,
	\nonumber\\&&\quad
	\frac{q^2}{\alpha_1}
        \left(\sqrt{\frac{K_L}{K_T}} + \sqrt{\frac{K_T}{K_L}}\right)
	e^{-|k|\xi_\parallel},
\label{eq:lindemann}
\end{eqnarray}
where $N_L$ is the number of layers, $k=0,2\pi/N_L,\ldots$ is expressed 
in units of the inverse layer spacing, $1/d$, and $q$ is in units of 
$1/\ell_\perp$. The constants $\alpha_1$, $\alpha_2$ and
$c_{66}$ are as in Ref.\ \cite{SHM2} and
\begin{eqnarray}
K_L(\vec q,k) &=&
    \frac{\alpha_2}{q^2} + \frac12\frac{\alpha_1}{q^2}t(1 - \cos(k))\\
K_T(\vec q,k) &=&
    c_{66}q^2 + \frac12\frac{\alpha_1}{q^2}t(1 - \cos(k)).
\end{eqnarray}
The Lindemann criterion predicts the quantum melting transition to
occur when the mean square displacement is\cite{RS}
\begin{eqnarray}
\langle u^2\rangle &\approx& \alpha_L^2 a_0^2,
\end{eqnarray}
where $\alpha_L\approx 0.1 - 0.2$ is a system-dependent ({\it a
priori\/} unknown) constant and $a_0$ is the lattice spacing.  We take
$\alpha_L=0.15$ and have plotted the resulting critical layer filling
factor $\nu_c^\prime(t)$ in Fig.~\ref{fig:lindemann}.

Our second approach to estimate the critical filling factor per layer
$\nu^\prime_c(t)$ is based on the depletion of the condensate \cite{SHM2}.  
This criterion is, perhaps, less direct in determining the melting point 
of the vortex lattice than the Lindemann criterion, and may tend to 
overestimate the critical filling factor $\nu^\prime_c(t)$. However, given 
its simplicity, it is valuable to know the results of the calculation 
based on this criterion.

\begin{figure}
\centerline{\epsfig{file=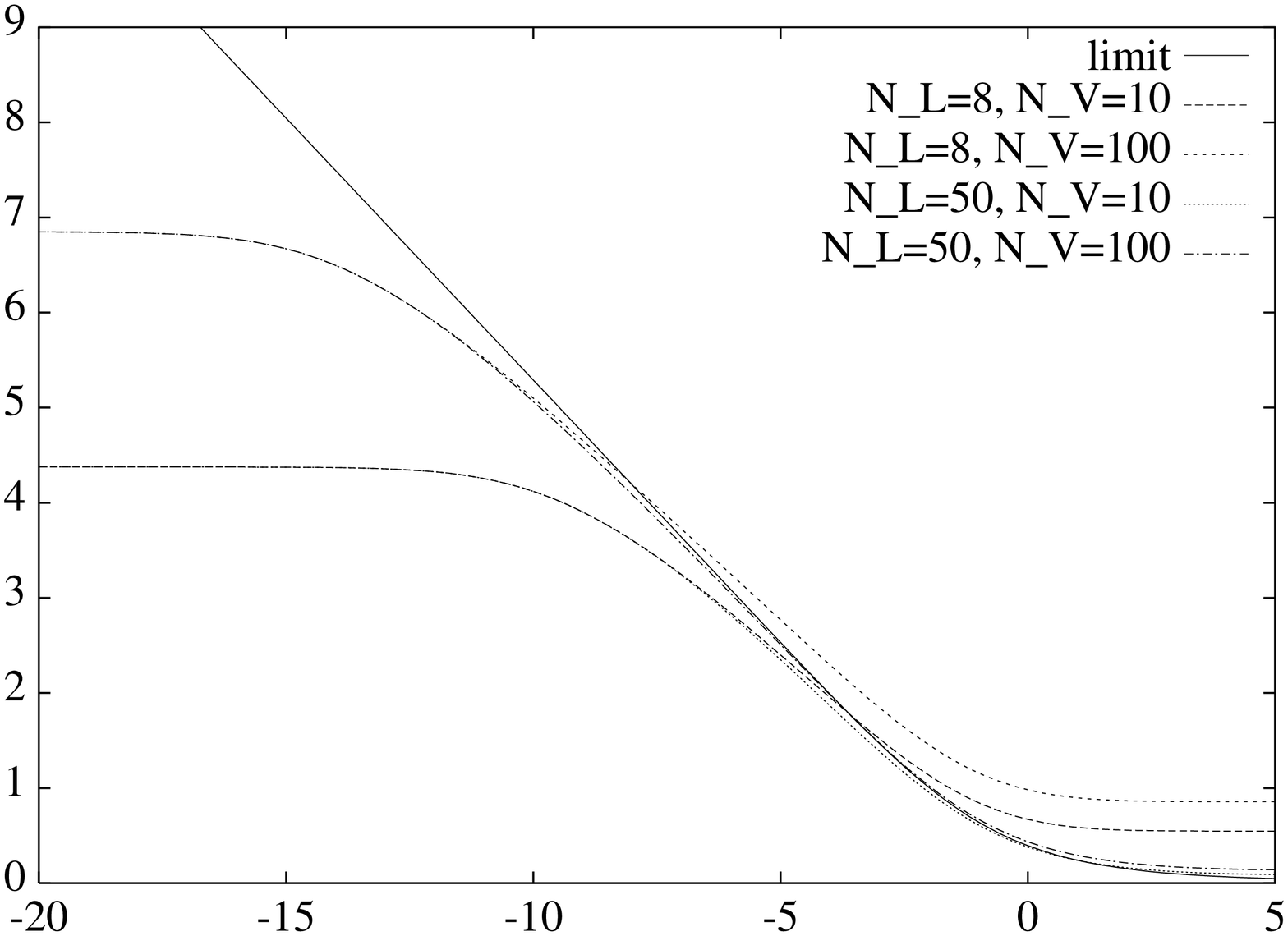,width=0.5\textwidth}}
\caption{Critical filling factor per layer $\nu_c^\prime(t)$ 
as a function of
$\ln(t/(\nu^\prime \gqH))$ for $N_L=8,\,50$ layers and $N_V=10,\,100$
vortices.The solid line indicates the limit $N_L\to\infty$, 
$N_V\to\infty$.
}
\label{fig:multimelt}
\end{figure}

The filling factor $\tilde{\nu}$ for bosons outside the
condensate is:
\begin{eqnarray}
\tilde{\nu} - \nu
            &=& \frac{\ell_\perp^2}{4\pi}\frac1{N_L}\sum_k
		\int_{\mbox{\footnotesize\it BZ}}\!d\vec q\,
		(\xi_{k,\vec q} / E_{k,\vec q} - 1),
\label{eq:nuc}
\end{eqnarray}
The one-layer expressions \cite{SHM2} ($\xi^0_{\vec q}$, 
$\lambda^0_{\vec q}$)
are generalized by $\xi_{k,\vec q} = \xi^0_{\vec q} + t (1 -
\cos(k))$, $E_{k,\vec q}=\sqrt{\xi_{k,\vec q}^2 -
|\lambda^0_{\vec q}|^2}$\@.  The $k=0$ integrand diverges when
$q\rightarrow 0$, due to phase fluctuations.  In a finite system
we can use the system size as an infrared cutoff,
$q_{\mbox{\footnotesize\it min}}=1/\sqrt{2\pi N_V}$.  We equate
the quantum melting point with the filling factor where half
the bosons are in an excited state and find the
critical filling factors shown in Fig.~\ref{fig:multimelt}.

The critical filling factor for large tunneling, determined in
this way, is well approximated by $\nu'_c(t\rightarrow\infty) =
\nu'_c(t\rightarrow0) / N_L$\@, in agreement with the intuition
that for large tunneling the system behaves as a single layer with 
total filling factor $\nu= N_L \nu'$.  As
can be observed in Fig.~\ref{fig:multimelt}, in a large range of
tunneling amplitudes the critical filling factor is independent of $N_L$ and
$N_V$.  Eq.~(\ref{eq:nuc}) can be approximated by a
converging integral, which is exact for $N_L\rightarrow\infty$,
$N_V\rightarrow\infty$\@.  For the parameters quoted above,
$N_L=50$, $N_V=100$, $\nu^\prime=1$, we find that the critical 
tunneling for quantum melting of the vortex lattice is $t_{c_1}\approx
0.18 \gqH$.

\subsection{Quantum phases of the multi-layer systems}
\label{sec:phases}
The melting of the vortex lattice leaves a multi-layer 
system of highly correlated quantum liquids. 
The nature of the quantum liquids is not {\em a priori\/}
clear. In the limit of vanishing tunneling, the isolated layers 
will display single-layer behavior.  The states formed close 
to the melting transition are, however, not necessarily 
adiabatically connected to these but could be separated by a 
quantum phase transition.

For $t$ smaller than but rather close to $t_{c_1}$ there is the 
possibility for correlated multi-layer states. For the case
of electronic quantum Hall systems, such correlated multi-layer
states have been proposed and analyzed (an early reference is
\cite{QJM}). For the atomic systems the multi-layer states can 
be very different, as there are no interlayer interactions
and interlayer correlations are driven by tunneling alone.

Our main interest is in the regime where the individual 
layers are weakly coupled. In this regime, the atoms will
be able to move in between layers and form an equilibrium 
density profile. In this regime we expect a local density 
approximation in both the coordinates $z$ and $r$ to be valid.
In section \ref{sec:expansion} below, we analyze this regime 
and work out the density profile arising after expansion. 

For $t\ll \gqH$, interlayer transport will stop and individual 
layers can be at a sharply quantized particle number, realizing 
what we call a quantum Hall-Mott (qH-Mott) phase. To determine the 
boundaries of these qH-Mott phases, we have analyzed a simple
effective Bose-Hubbard model. In the simplest case we take
quadratic in-plane confinement and assume that all layers form 
a Laughlin state (with filling factor per layer $\nu'=1/2$). 
For this situation we obtained 
the following expression for the critical tunneling $t$ for 
entering the qH-Mott state with $M$ particles, 
\be
t_{c_2}(\mu) = -\lambda_2
  [M + {1 \over 2} - {\mu \over 2 \lambda_2}]
  [M - {1 \over 2} - {\mu \over 2\lambda_2}] \ ,
\ee
for a chemical potential in the interval $M-{1 \over 2}
\leq {\mu \over 2 \lambda_2} \leq M+{1 \over 2}$. 
As in the standard Bose-Hubbard model, the curves $t_c(\mu)$ 
define characteristic `lobes' in the $\mu$-$t$ 
plane, marking the boundaries of the qH-Mott phases. We observe
that $t_{c_2}(\mu)$ has a maximum of $\lambda_2/4$, which is of 
order $\gqH/M$. This scale is in agreement with the fact that
the lowest excitations of the quantum Hall droplets are
edge excitations with energies of order $\gqH/M$.

\begin{figure}
\epsfig{file=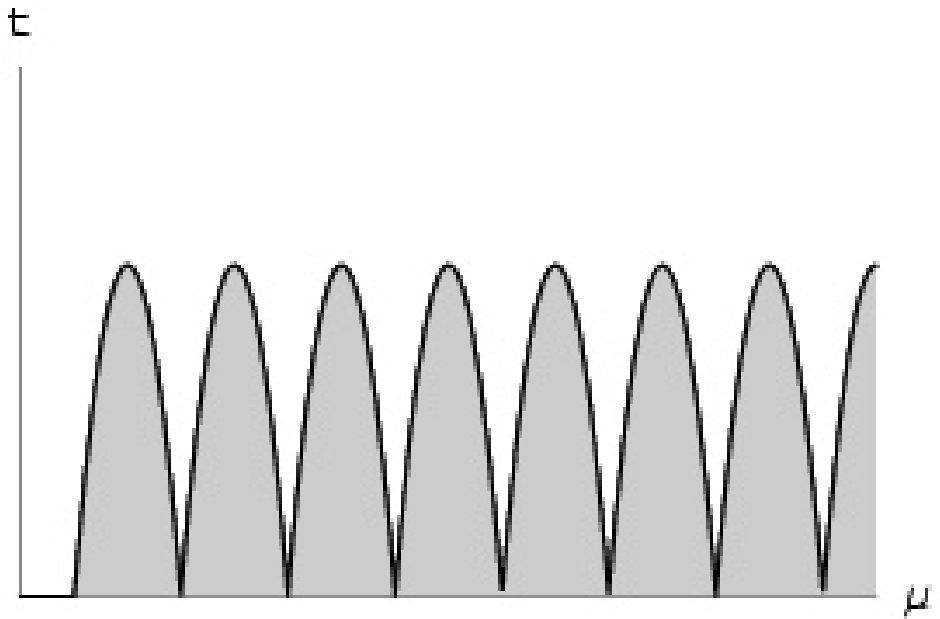,width=65mm} 
\vspace{3mm}
\epsfig{file=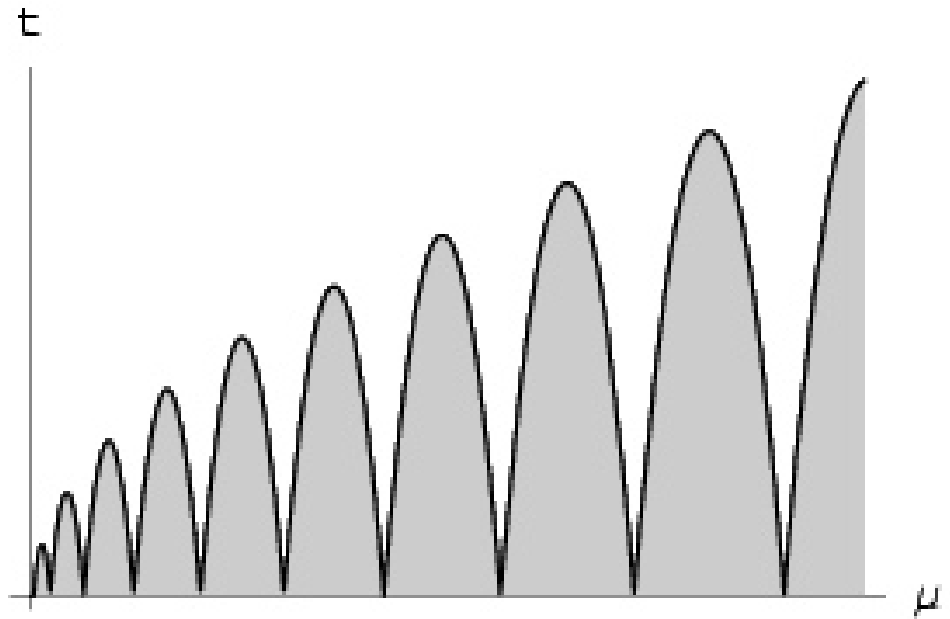,width=65mm} 
\caption{Phase diagrams in the $\mu(z)$-$t$ plane, for
a situation with $\nu=1/2$ Laughlin quantum Hall states in
the layers. The shaded areas correspond to quantum Hall Mott 
states, where the particle number per layer is sharply quantized. 
In the remaining part of the diagram there are interlayer 
fluctuations, giving rise to an axial superfluid. The top and 
bottom panels are for quadratic and quartic confinement, 
respectively.}

\label{fig:mu-t}
\end{figure}

In Fig.~\ref{fig:mu-t} we display $\mu(z)$-$t$ diagrams for
Laughlin states in quadratic and quartic in-plane confinement.
In the quartic case, the maximum of $t_{c_2}$ grows with $\mu$.
This makes possible a situation where the central layers of the
stack (near $z=0$) display qH-Mott states, while the outer layers
are in the fluctuating regime.

We wish to emphasize that these quantum Hall-Mott phases are
related to the well known Mott phase.  When a 1$D$ lattice of atoms
in the Mott phase is subjected to rotation (with fixed $M$), each site
will expand to a layer with a well defined filling factor $\nu'$\@.
When $\nu'\approx\nu_c$, the single layer critical filling factor, the
system will undergo a quantum phase transition to the quantum
Hall-Mott phase.  This transition occurs when $t<t_{c_2}$, where we
extend the definition of $t_{c_2}$ to the Mott phase.

An interesting point in the $t-\nu'$ phase diagram (for fixed $M$)
is $(t, \nu') = (t^*, \nu^*)$, where $t_{c_1}=t_{c_2}\equiv t^*$ and
$\nu^*$ is the critical filling factor for $t=t^*$\@.  At this point,
the Mott transition coincides with the quantum melting transition.

\subsection{Density profile after expansion}
\label{sec:expansion}
In this section we focus on the regime where the tunneling
is such that we have a stack of weakly coupled quantum Hall 
layers. We can apply the ideas developed in section
\ref{sec:single}, where we showed how each individual layer 
is built from local pieces of quantum Hall fluid.
Having a slowly varying chemical potential $\mu(z)$ implies
that the density landscape varies slowly from layer to layer,
and that processes where atoms tunnel from the edge of a local
quantum Hall fluid in one layer $i$ to the corresponding edge in
an adjacent layer $i\pm 1$, are possible. The energy scale for
such inter-layer edge excitations is of order $\gqH/N_i$,
with $N_i$ the number of particles in layer $i$. We assume that the 
tunneling $t$ is large on this scale, such that tunneling events 
can establish an equilibrium density 
profile $n(r,z)$. 

The density profile can be written as
\be
  n(r,z) = n_0 \nu[\mu_0 - \mu_2(z/d)^2 - V(r)] \ ,
\ee
with $\nu[\mu]=\nu_i$ for $\mu_c^{(i)}<\mu<\mu_c^{(i+1)}$.
A measurement of particle number per layer averages this over $r$:
\be
  \rho(z) = \int 2 \pi dr \, n(r,z) \ .
\ee
The profiles $\rho(z)$ have been discussed in \cite{letterCvLRS}.
Here we focus instead on the density profile arising after switching 
off the optical lattice potential in the $z$-direction and the trapping 
potential. In this expansion,
each tightly confined layer expands enormously in the $z$ direction,
until all layers overlap. The expansion image is thus obtained by 
averaging over $z$,
\be
  \rho(r) = \int dz \, n(r,z) \ .
\label{eq:rho-r}
\ee
Expansion in the radial direction, which is needed for
an actual measurement of $\rho(r)$, will not affect the density
profile other than by an overall expanding scale\cite{RC}.

\begin{figure}
\epsfig{file=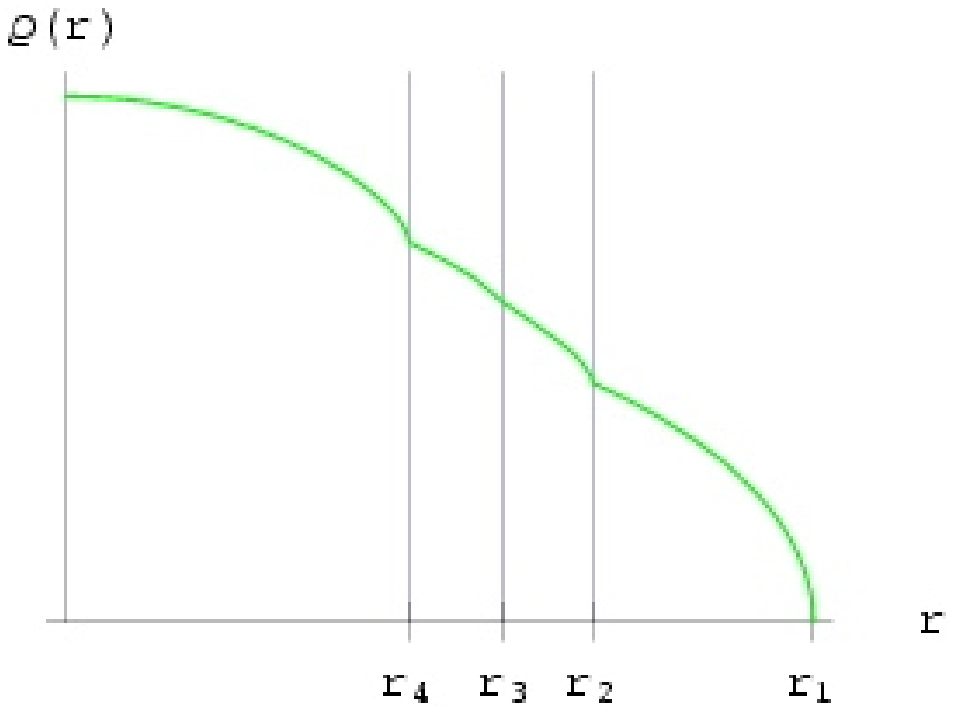,width=70mm}
\vskip 4mm
\epsfig{file=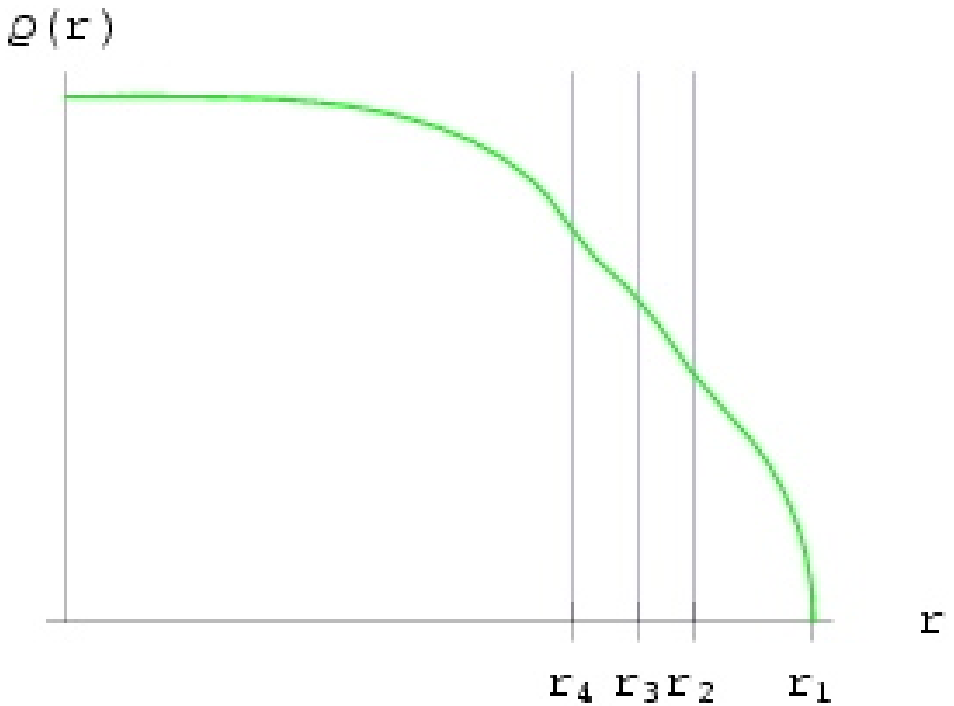,width=70mm}
\caption{Radial density profile $\rho(r)$, for a
system with quantum Hall liquids at $\nu_1=1/2$, $\nu_2=2/3$, 
$\nu_3=3/4$ and $\nu_4=1$. These profiles arise after expansion 
in the $z$-direction. The top panel is for quadratic in-plane 
confinement, with parameters chosen such that the landscape in the
central layer (at $z=0$) coincides with the top curve in 
Fig.~\ref{fig:steps2}. The bottom panel assumes quartic 
confinement.}
\label{fig:profile1}
\end{figure}

\begin{figure}
\vspace{5mm}
\epsfig{file=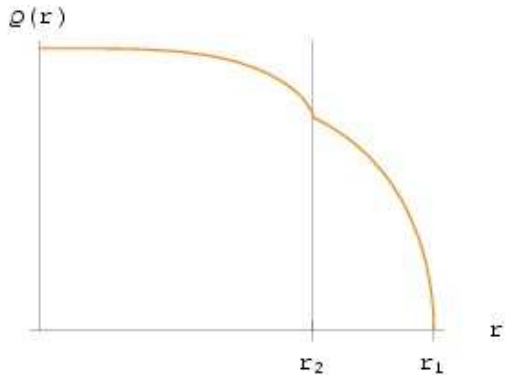,width=70mm}
\caption{Radial density profile $\rho(r)$, for a
system with quantum Hall liquids at $\nu_1=1/2$ and $\nu_2=2/3$,
for quartic in-plane confinement.}
\label{fig:profile2}
\end{figure}

The integral in Eq.~\ref{eq:rho-r} is easily worked out. If we assume 
quadratic in-plane
confinement, $V(r)={1 \over 2} k_2 r^2$, the profile takes the form of
a sum of semi-circle arcs
\be
\label{eq:arcs}
\rho(r) = n_0 \left( {2 k_2 \over \mu_2} \right)^{1/2}
\sum_i [\nu_i-\nu_{i-1}] \sqrt{r_i^2 - r^2}
\ee
with
\be
r_i = \left( {2 (\mu_0-\mu_c^{(i)}) \over k_2} \right)^{1/2} \ .
\ee
For quartic confinement the result is
\be
\rho(r) = n_0 \left( {k_4 \over \mu_2} \right)^{1/2}
\sum_i [\nu_i-\nu_{i-1}] \sqrt{r_i^4 - r^4}
\ee
with
\be
r_i = \left( {4 (\mu_0-\mu_c^{(i)}) \over k_4} \right)^{1/4} \ .
\ee
It is clear that, while the expansion does wash out the sharp
steps in the density profiles, it leaves cusps in the averaged
profiles. For in-plane confinement $V(r)\propto r^\alpha$
these cusps take the form of square root singularities,
$\rho(r)\propto \sqrt{R^\alpha - r^\alpha}$. In Figs.\
\ref{fig:profile1} and \ref{fig:profile2} we display
some of the profiles $\rho(r)$.
The positions of the cusps provide information on the incompressible
liquid states. Fitting the measured density profile to (\ref{eq:arcs}), involving the
set of filling factors $\nu_i$, allows the critical chemical potentials,
$\mu_c^i$, to be measured. This would provide direct information on the
energetics and incompressibility of the correlated quantum liquids.

We now estimate the magnitude of fluctuations that affect these
mean field density profiles. Focusing on a qH fluid ($\nu_k$) 
in one of the layers, we can evaluate the energy for a 
displacement of the outer edge, $r_k \to r_k + \delta r_k$.
In evaluating this energy, we assume that the chemical potential
at the center of the layer is of order $\gqH$ (which is the
natural scale in the present regime). With that we find
\be
\delta E \simeq \gqH \left( {\delta r_k \over \ell_\perp} \right)^2
\ee
Equating this to the scale set by the tunneling gives the 
estimate
\be
\delta r_k \simeq \ell_\perp \sqrt{{t \over \gqH}} \ .
\ee
We conclude that, for $t$ of order $\gqH$ or smaller, the
fluctuations induced on $r_k$ are of order $\ell_\perp$ or less.
This means that, in the presence of tunneling, the steps
within each of the layers remain well-defined on a scale
$\ell_\perp$. Accordingly, one expects that the cusps in the 
profile after expansion will be recognizable if the overall
extent (in the $x$-$y$ plane) of the original cloud is large 
on the scale of $\ell_\perp$. 

The features in $\rho(r)$ can be sharpened further by eliminating 
some of the outer layers from the stack, keeping only a number of 
layers centered around $z=0$, or by having an axial confinement
$\mu(z)$ that is steeper than quadratic. 

\vskip 3mm

We thank J.~Dalibard and V.~Schweikhard for illuminating discussions
on the prospects for experimental realization of atomic quantum Hall
states. This research was supported by the Netherlands Organisation for
Scientific Research, NWO, the Foundation FOM of the Netherlands
(JWR, FJMvL and KS), and by EPSRC grant GR/S61263/01 (NRC).

\end{document}